\documentclass[cernpreprint,texlive=2011,txfonts,UKenglish,texmf]{atlasdoc}
\pdfoutput=1

\usepackage{atlaspackage}
\usepackage{atlasbiblatex}
\usepackage{atlasphysics}

\AtlasTitle{Determination of the ratio of $b$-quark fragmentation fractions $f_s/f_d$ in \textit{pp} collisions at $\sqrt{s}=7\ \mathrm{TeV}$ with the ATLAS detector}

\author{The ATLAS Collaboration}

\AtlasAbstract{
  With an integrated luminosity of $2.47~\textrm{fb}^{-1}$ recorded by the ATLAS experiment at the LHC, the exclusive decays $\Bs\rightarrow \Jpsi\phi$ and $\Bd\rightarrow \Jpsi K^{*0}$ of $B$ mesons produced in $pp$ collisions at $\sqrt{s}=7\ \mathrm{TeV}$ are used to determine the ratio of fragmentation fractions $f_s/f_d$. From the observed $\Bs\rightarrow \Jpsi\phi$ and $\Bd\rightarrow \Jpsi K^{*0}$ yields, the quantity $\frac{f_s}{f_d}\frac{\mathcal{B}(\Bs\rightarrow \Jpsi \phi)}{\mathcal{B}(\Bd\rightarrow \Jpsi K^{*0})}$ is measured to be $0.199 \pm 0.004(\mathrm{stat}) \pm 0.008(\mathrm{sys})$.  Using a recent theory prediction for $\frac{\mathcal{B}(\Bs\rightarrow \Jpsi \phi)}{\mathcal{B}(\Bd\rightarrow \Jpsi K^{*0})}$ yields $\frac{f_s}{f_d} = 0.240 \pm 0.004(\mathrm{stat}) \pm 0.010(\mathrm{sys}) \pm 0.017(\mathrm{th})$. This result is based on a new approach that provides a significant improvement of the world average. 
}

\AtlasJournal{Phys.\ Rev.\ Lett.}
\PreprintIdNumber{CERN-PH-EP-2015-165}
\AtlasDate{}
\AtlasJournalRef{}
\AtlasDOI{}
\arXivId{}

\begin{document}

\maketitle

The production rate of $\Bs$ ($\Bd$) mesons is a product of the $b\bar{b}$ cross section, the instantaneous luminosity and the probability that the $\bar b$-quark is bound to an $s$-($d$-) quark. The latter, denoted by the fragmentation fraction $f_s$ ($f_d$), depends on the probability that in pQCD-inspired calculations~\cite{pQCD1,pQCD2} a soft gluon splits into $s\bar{s}$ ($d\bar{d}$) and that the overlap of the $\bar b$ and  $s~(d)$ wave functions is sufficiently large to produce a $\Bs$ ($\Bd$) bound state. In a similar fashion, $B^+$ mesons, $B_c$ mesons and $b$-baryons are produced at the LHC with respective fragmentation fractions $f_u$, $f_c$ and $f_{\mathrm{baryon}}$. The fragmentation fractions are about $40\%$ each for $ u$- and $ d$-quarks, $10\%$ for $s$-quarks, at the percent level for $ c$-quarks and $\sim 8\%$ for baryon production satisfying the constraint $f_u+f_d+f_s+f_c+f_{\mathrm{baryon}}=1$.
Precise knowledge of the fragmentation fractions is essential for measuring $b$-hadron cross sections and branching fractions at the LHC. In particular for rare decays, such as the branching fraction measurement of $\Bs\rightarrow\mu^{+}\mu^{-}$~\cite{ATLAS_mumu, LHCb_mumu, CMS_mumu}, a precise knowledge of $f_s/f_d$ is important since it improves the sensitivity of searches for new physics processes beyond the Standard Model (SM). The fragmentation ratio $f_s/f_d$ is a universal quantity that was measured by LEP experiments~\cite{HFAG}, CDF~\cite{CDF} and LHCb~\cite{LHCB_fsfd, LHCbUpdated}. This Letter presents a measurement of $f_s/f_d$ using  ${\Bs \rightarrow \Jpsi\phi}$ and ${\Bd \rightarrow \Jpsi K^{*0}}$ decays. 

The ratio of fragmentation fractions $f_s/f_d$ is extracted from the measured $\Bs \rightarrow \Jpsi \phi$ and $\Bd \rightarrow \Jpsi K^{*0}$ signal yields, $N_{\Bs}$ and $N_{\Bd}$. These are converted into $\Bs$ and $\Bd$ meson yields after dividing by the branching fractions of the relevant decays and correcting for the relative efficiency $\mathcal{R}_{\mathrm{eff}}$ that is expressed as a product of acceptance and selection efficiency ratios for the two modes and is determined from  Monte Carlo (MC) simulations:
 
\begin{equation}
\frac{f_{s}}{f_{d}}=\frac{N_{\Bs}}{N_{\Bd}} 
\frac{\mathcal{B}(\Bd\rightarrow \Jpsi K^{*0})}{\mathcal{B}(\Bs\rightarrow \Jpsi \phi)} 
\frac{\mathcal{B}( K^{*0}\rightarrow K^{+}\pi^{-})}{\mathcal{B}(\phi \rightarrow K^{+}K^{-})} \mathcal{R}_{\mathrm{eff}},
\label{eq:fsfd}    
\end{equation}  

\noindent where the $\Jpsi$, $\phi$ and $K^{*0}$ are reconstructed in their 
$\Jpsi \rightarrow \mu^+ \mu^-$, $\phi \rightarrow K^+ K^-$ and $K^{*0} \rightarrow K^+ \pi^-$ final states~\cite{cc}, respectively. The data sample consists of $pp$ collisions collected with the ATLAS detector at $\rts=7~\textrm{TeV}$ corresponding to an integrated luminosity of $2.47 \pm 0.04~\textrm{fb}^{-1}$. The ATLAS multipurpose detector is described in detail in Ref.~\cite{ATLAS_experiment}.

The {\sc PYTHIA} 6 and 8~\cite{PYTHIA6,PYTHIA8} MC generators with parameters tuned to reproduce ATLAS data~\cite{ATLAStune} are used to simulate background and signal events, respectively. For the signal channels, the angular distributions are produced with the measured polarization parameters~\cite{PDG}. The detector response for the generated events is simulated with {\sc Geant4}~\cite{Geant4, ATLASsim}.  

The $\Bs \rightarrow \Jpsi\phi$ and $\Bd \rightarrow \Jpsi K^{*0}$ signal candidates consist of two muons and two hadrons originating from a common secondary vertex. The $\Jpsi$ candidates are selected from the dimuon trigger sample requiring two oppositely charged muon candidates, each having a transverse momentum of $\pt>$ 4 GeV. Reconstructed muon candidates are categorized either as combined or segment-tagged muons. A combined muon consists of an inner detector (ID) track combined with a muon spectrometer (MS) track using tight matching criteria, while a segment-tagged muon requires an ID track and track segments in the MS that are not reconstructed as an MS track~\cite{ATLAS_experiment}. The two muons, of which at least one must be a combined muon, are fitted to originate from the same two-track vertex. The vertex fit chi-square per degree of freedom ($\mathrm{dof}$) is required to be $\chi^{2} / \mathrm{dof}<10$. To improve the sample purity, each muon track must have at least one hit in the pixel detector, more than five hits in the silicon strip detector and at least one hit in the transition radiation tracker that reduces the pseudorapidity coverage to $|\eta|<2.0$~\cite {ATLAS_coordinate}.

Since the dimuon mass resolution is different for muons reconstructed in the endcaps ($1.05 < |\eta| < 2.5$) and for muons reconstructed in the barrel  ($|\eta| < 1.05$), all accepted $\Jpsi$ candidates are divided into three classes: two barrel muons (BB), one endcap and one barrel muon (EB), and two endcap muons (EE). The parameters describing the dimuon mass distribution in the $\Jpsi$ signal region for the three pseudorapidity classes in data and in $\Bs \rightarrow \Jpsi\phi$ and $\Bd \rightarrow \Jpsi K^{*0}$ MC signal samples are extracted from maximum-likelihood fits. Signal events are selected requiring mass windows of $\pm 3 \sigma$ around the $\Jpsi$ peak in data and simulations. For data,  the selected signal regions  are 2.991--3.197 GeV for BB, 2.955--3.235 GeV for EB and 2.914--3.275 GeV for EE classes, while in simulations they are slightly smaller.

The $\Bs$ candidates are reconstructed from a $\Jpsi$ candidate plus two oppositely-charged hadrons with a kaon mass hypothesis assigned. The dimuon mass is constrained to the $\Jpsi$ mass~\cite{PDG} and the $\Jpsi$  and two kaons have to originate from the same vertex. All combinations are accepted if $\pt (\Bs) > 8$ GeV, $\chi^{2} / \mathrm{dof} < 3$ for the vertex fit and the  $K^+ K^-$ invariant mass lies in the range determined by $\pm 2$ natural widths ($\Gamma_\phi)$ around the $\phi$ mass peak, $1011 < m_{K^{+}K^{-}} < 1028$~MeV. The $m_{K^{+}K^{-}}$ distribution is modeled with a Breit-Wigner line shape convolved with a Crystal Ball function~\cite{CB}. The selected mass window retains 85\% of signal events.

The $\Bd$ candidates are reconstructed in a similar way. Here, one track of the $K^{*0}$ decay is assigned a kaon mass hypothesis and the other track a pion mass hypothesis. Since ATLAS has limited kaon-pion separation capability in the momentum range relevant for this analysis, both $K \pi$ mass assignment combinations are tested. That with mass closest to the nominal $K^{*0}$ mass is chosen yielding the correct $K \pi$  selection for 86\% of all $K^{*0}$ candidates. The probability density function (PDF) for the invariant mass of correctly selected $K \pi$ candidates is modeled with a relativistic Breit-Wigner line shape convolved with a Crystal Ball function, while that where the $K$ and $\pi$ are swapped is modeled with a Gaussian function. The decay $\Bs \rightarrow \Jpsi\phi$ produces a peaking background in $\Bd \rightarrow \Jpsi K^{*0}$ that appears in the low $K^{*0}$ mass region. To remove this contribution, the selected $K^{*0}$ region is constrained to one $K^{*0}$ decay width around the $K^{*0}$ mass peak, corresponding to $ 847 < m_{K \pi}< 942$ MeV for data. Since the $K^{*0}$ line shape is narrower in the MC simulations than in data, the $K\pi$ mass selection needs to be adjusted in simulations to produce identical efficiencies in data and simulations. For the $K^+ K^-$ mass selection, a similar procedure is used.
 
The signal-to-background ratios for $\Bs \rightarrow \Jpsi\phi$ and $\Bd \rightarrow \Jpsi K^{*0}$ decays are optimized using three variables with high background suppression power: the $\chi^{2} / \mathrm{dof}$ of the $B$ vertex fit, the transverse decay length $L_{xy}$ defined as the length of the vector from the primary vertex (PV)~\cite{PV} to the $B$ decay vertex in the transverse plane, and the pointing angle $\alpha$ defined as the angle between the $B$ meson transverse momentum and $L_{xy}$. If more than one PV candidate exists, the one is selected for which
 the sum of squared transverse momenta of all tracks originating from the vertex, $\sum \pt^{2}$, yields the highest value. The $\chi^{2} / \mathrm{dof}$, $L_{xy}$ and $\alpha$ selection criteria are optimized using simulated $\Bs \rightarrow \Jpsi\phi$ and $\Bd \rightarrow \Jpsi K^{*0}$ events for signal and data sidebands for background. 

To produce similar $\pt$ and $\eta$ distributions in data and MC, data-driven weights are obtained by the following procedure. Sideband-subtracted $\Bs \rightarrow \Jpsi\phi$ ($\Bd \rightarrow \Jpsi K^{*0}$) $\pt$ and $\eta$ distributions from data are compared with corresponding distributions in simulation in the signal region, $ 5.32 < m_{J/\psi \phi}  < 5.42~(5.21 < m_{J/\psi K^{*0}} < 5.35)~\rm GeV$. The upper and lower sidebands $5.20< m_{J/\psi \phi}  <  5.25~(5.09 < m_{J/\psi K^{*0}} <  5.16)~\rm GeV$ and $5.48 < m_{J/\psi \phi}  <  5.53~(5.40 < m_{J/\psi K^{*0}} <  5.47)~\rm  GeV$ are selected such that their summed yields represent the expected backgrounds in the signal region for the data. The weights are obtained by dividing the yield in each $\pt$ and $\eta$ bin in data by the corresponding yield of the MC sample using only events with odd event numbers. Thus, for each bin ($i$) and ($j$) of the $\pt$ and $\eta$ distributions, a weight is determined as a product of a $\pt$-dependent and $\eta$-dependent weights:

 \begin{equation} 
 W_{ij}(\pt, \eta)=\dfrac{n^{\mathrm{data}}_{i}(\pt)}{n^{\mathrm{MC}}_{i} (\pt)} \times \dfrac{n^{\mathrm{data}}_{j}(\eta)}{n^{\mathrm{MC}}_{j} (\eta)},
 \end{equation}
 
\noindent where $n^{\mathrm{data/MC}}_{i}(\pt)$  is the normalized number of entries in the $\pt$ bin $i$ and $n^{\mathrm{data/MC}}_{j}(\eta)$ is that in the $\eta$ bin $j$. To obtain good agreement between data and simulation, the procedure is repeated twice. The two sets of weights are multiplied and are used to correct the $\pt$ and $\eta$ distributions of the MC sample with even event numbers. From the corrected MC samples, distributions for  $\chi^{2} / \mathrm{dof}$, $L_{xy}$ and $\alpha$ are determined that are in good agreement with those measured in the data. The correlation between $\pt$ and $\eta$ is small and is accounted for in the systematic error. 

For both modes, the dominant background originates from a $\Jpsi$ produced at the PV plus two oppositely charged hadrons (direct  $\Jpsi$)~\cite{ATLAS_pp}. Since the hadrons are not associated with any $\Bs~(\Bd)$ decay, the $\Jpsi K^+ K^-~(\Jpsi K^+ \pi^-)$ invariant-mass spectrum does not peak but decreases with mass. Another large background consists of two random low-momentum, oppositely charged muons combined with two random charged hadrons. Here, the dimuon mass distribution does not peak at the $\Jpsi$  nor  does the four-particle mass show any peaking structure. Inclusive decays $B \rightarrow \Jpsi X$ where $X$ is a single hadron or a collection of hadrons provide a source of background that is very similar to the signal. If $X$ consists of exactly two charged-particle tracks (without any $\pi^0$), the mode is topologically indistinguishable from the signal mode. Self-cross-feed in which one or both hadrons from the $\phi~(K^{*0})$ decay are replaced with random hadrons is negligible. In addition, peaking backgrounds from $\Bd \rightarrow \Jpsi K^{*0}$ and $\Bd \rightarrow \Jpsi K^{+}\pi^{-}$ contribute to $\Bs \rightarrow \Jpsi \phi$  while 
$\Bd \rightarrow \Jpsi K^{+}\pi^{-}$ also contributes to $\Bd \rightarrow \Jpsi K^{*0}$.

To reduce these backgrounds, the $\chi^{2} / \mathrm{dof}$, $L_{xy}$ and $\alpha$ selections are optimized for each mode separately by determining the maximum value of $S/\sqrt{S+B}$ as a function of selected values for the observable to be optimized, where $S$ represents the signal yield obtained from simulation and $B$ is the background extracted from data sidebands. For the $\Bs$ ($\Bd$) mode, the optimization yields $\chi^{2} / \mathrm{dof} <  2.4~(2.6)$, $L_{xy}  > 0.26~(0.30)$ mm and  $\alpha < 0.14~(0.12)$~rad. In combination with the $\Jpsi$ mass requirement, the $\chi^{2} / \mathrm{dof}$ selection reduces the combinatorial background significantly, while the $L_{xy}$ and $\alpha$ selections remove most of the direct $\Jpsi$ background. 

In the final sample, the signal yields $N_{\Bs}$ and $N_{\Bd}$ are extracted from unbinned extended maximum-likelihood fits to the $\Jpsi K^+ K^-$ and $\Jpsi K^+ \pi^-$ invariant-mass spectra, respectively. The $\Bs$ signal PDF is modeled with three Gaussian functions with common mean that is determined from the fit while widths and fractions are fixed to the values obtained from MC simulations. To account for possible width differences in the two narrowest Gaussian functions between data and simulation, an additional scale factor is introduced, which is left free in the fit. The peaking background PDF is modeled with a Crystal Ball function with parameters fixed to the values obtained in simulations. The peaking background yield of $652\pm 93$ events is calculated from the $\Bd$ signal yield. The selection efficiencies of both peaking background modes are determined from simulation and are fixed in the fit to data. The remaining residual backgrounds are modeled with an exponential function leaving fraction and exponent free in the fit to data.

The $\Bd$ signal PDF is parametrized with three Gaussian functions that describe both the correctly reconstructed and swapped $K^{+}\pi^{-}$ events. The PDF of the peaking background is modeled with a sum of Crystal Ball and Gaussian functions for which the relative $\Bd \rightarrow \Jpsi K^+ \pi^-$ yield with respect to that of the $\Bd \rightarrow \Jpsi K^{*0}$ signal is determined from the corresponding branching fractions and selection efficiencies, yielding $(4.7\pm 2.4)\%$. Most of the residual background is modeled with an exponential function while partially reconstructed $B\rightarrow\Jpsi X$ decays require parameterization with a complementary error function. All parameters of the residual background PDFs are left free in the fit.

\begin{figure}
\centering
\includegraphics[width=3in]{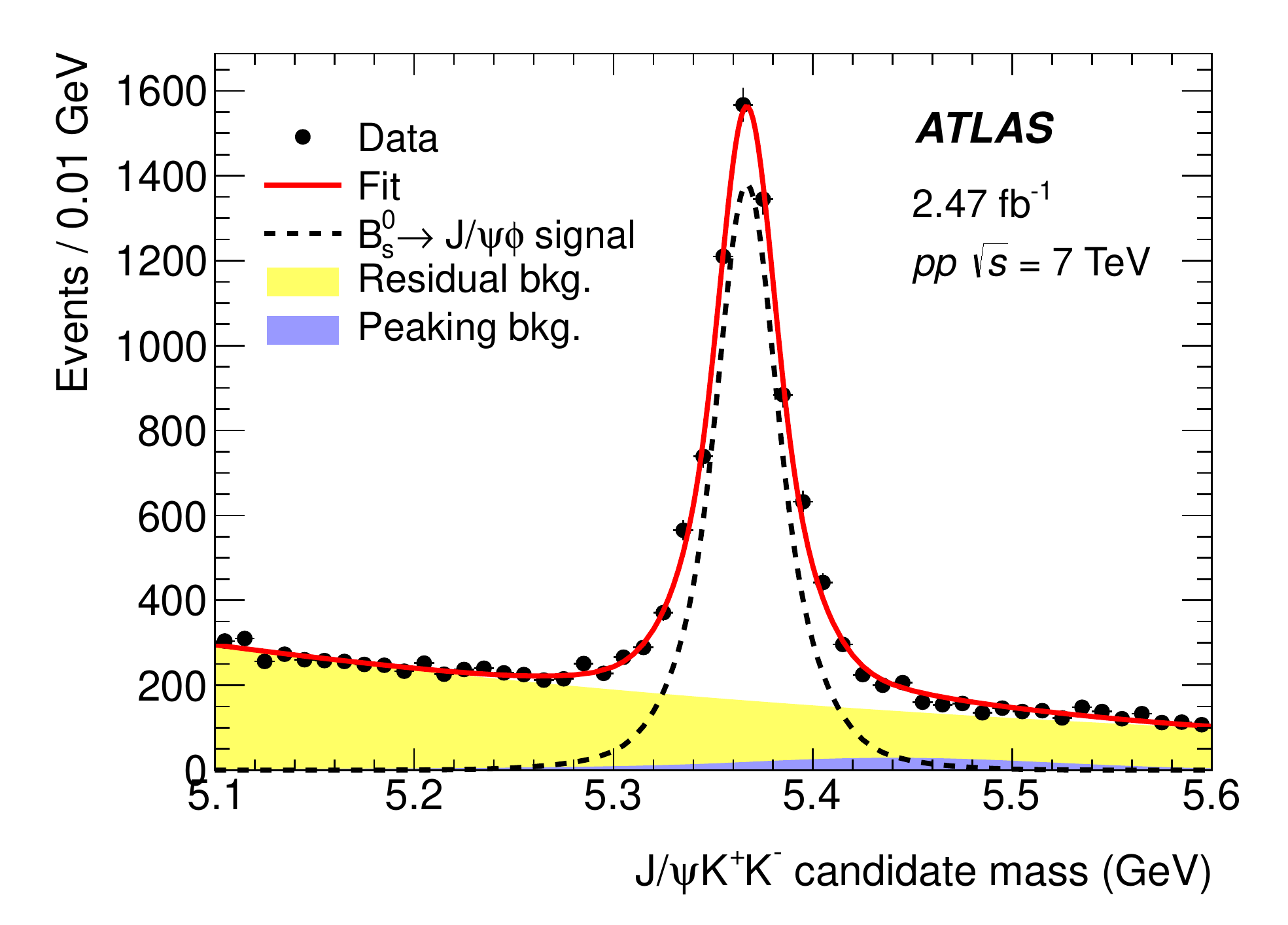}
\includegraphics[width=3in]{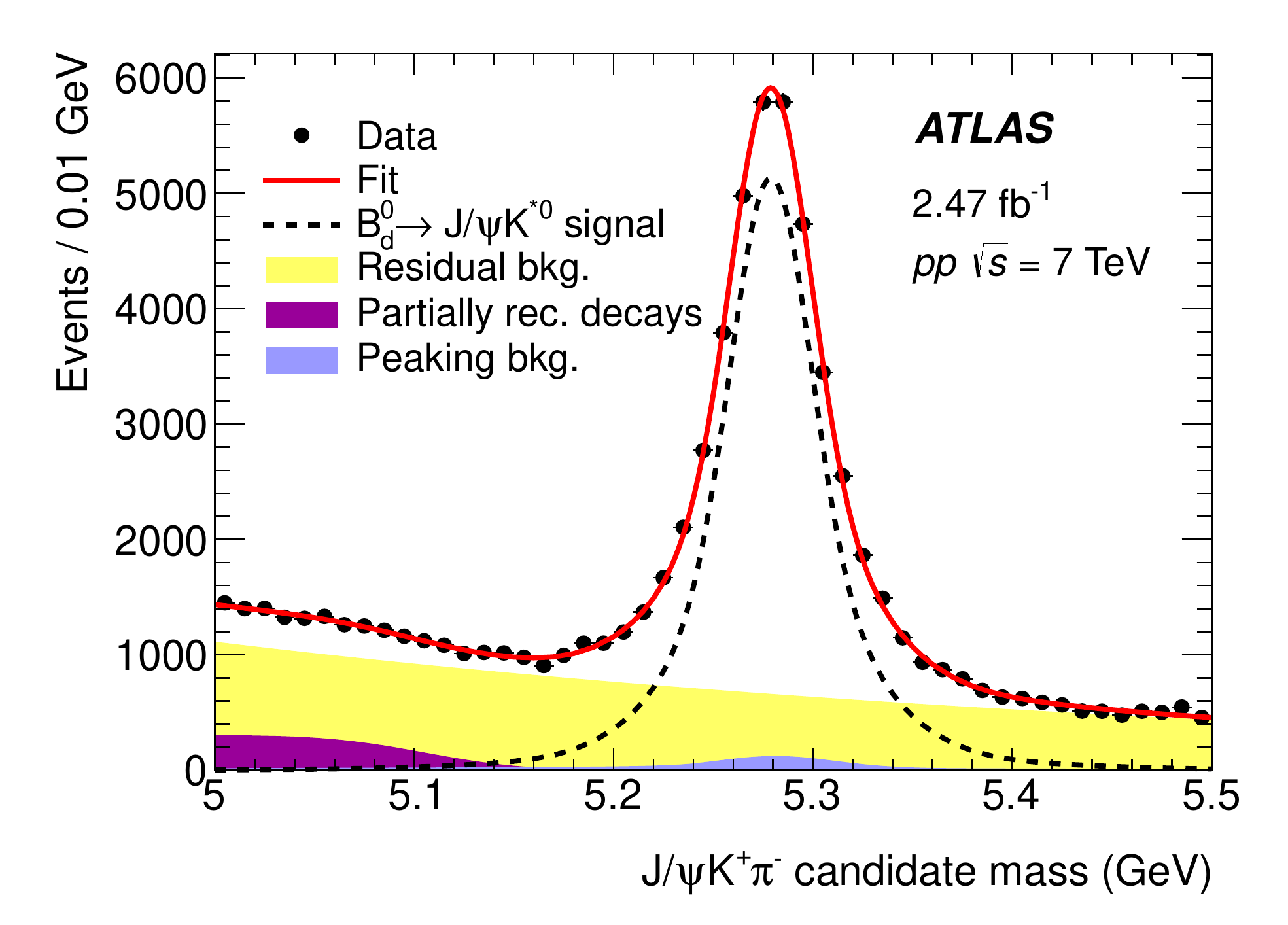}
\caption{The invariant-mass spectra of $\Bs \rightarrow \Jpsi \phi$ (left) and $\Bd \rightarrow \Jpsi K^{*0}$ decays (right) for data (points with error bars), total fit (solid), signal (dashed), residual background (yellow shaded), partially reconstructed events (magenta shaded) and peaking background (blue shaded).}
\label{fig:Bfits}
\end{figure} 

Figure~\ref{fig:Bfits} shows the measured $\Jpsi \phi$ and $\Jpsi K^{*0}$ invariant-mass spectra with fits overlaid. The fits yield $N_{B^0_s} = 6640 \pm 100$ $\Bs \rightarrow \Jpsi \phi$ and $N_{B^0_d}=36290 \pm 320$ $\Bd \rightarrow \Jpsi K^{*0}$ signal events. The $\chi^{2} / \mathrm{dof} $ values of the fits are 0.959 for $\Bs$ and 0.945 for $\Bd$ indicating that both fits describe the data well.

The additive systematic uncertainties result from the $\Bs \rightarrow \Jpsi \phi$ and $\Bd \rightarrow \Jpsi K^{*0}$ signal and  background parameterizations. The contribution from the signal shape parameterization is calculated by varying the five fixed parameters within $\pm 1\sigma$ in a multivariate Gaussian function that takes into account all correlations.  For non-peaking backgrounds, the exponential function is replaced with a second-order polynomial for the $\Bs$  and  with a  second-order polynomial plus an error function for the $\Bd$. The difference in signal yield with respect to the nominal fit is taken as a systematic error. For peaking backgrounds, the fixed parameters are varied by $\pm 1 \sigma$ and the difference with respect to the nominal yield is taken as a systematic error. In addition, since S-wave contributions from $\Bs \rightarrow \Jpsi K^{+}K^{-}$ and $\Bs \rightarrow \Jpsi f_0(\mathrm{980})$ decays to $\Bs \rightarrow \Jpsi \phi$ and $\Bd \rightarrow \Jpsi K^{*0}$ are neglected in the fits, an uncertainty is derived using the ATLAS measured contribution of 2.4\%~\cite{AtlasAngular} for $\Bs \rightarrow \Jpsi \phi$, and the contribution of 1\% for $\Bd \rightarrow \Jpsi K^{*0}$ derived from the MC simulation. All additive systematic errors are added in quadrature, yielding total additive uncertainties of 220 $N_{B_s^0}$ and 650 $N_{B_d^0}$ events. 

The multiplicative systematic uncertainty includes contributions from the relative efficiency and the branching fractions of the $\phi$ and $K^{*0}$ decays. The uncertainty on the relative efficiency is dominated  by the uncertainty on the $\phi/K^{*0}$ selection (1.2\%) which is obtained by varying the fixed fit parameters in the $\phi$ and $K^{*0}$ fits by $\pm1 \sigma$ and adding all contributions in quadrature. Other uncertainties from the $\Jpsi$ selection (0.2\%), reweighting (0.4\%), $\Bs$ and $\Bd$ lifetimes (0.002\%) and the contribution due to uncertainties in the  polarization parameters (0.01\%) are negligible. Varying the selection criteria of $\chi^{2} / \mathrm{dof}$, $L_{xy}$, and $\alpha$ gives negligible contributions. Table~\ref{tab:systematic} summarizes the contributions of the additive and multiplicative systematic errors. 

\begin{table}
\centering
\caption{Measured $B_s^0$ and $B_d^0$ signal yields, the efficiency ratio
 $\mathcal{R}_{\mathrm{eff}}$ extracted from simulations, world averages for
$\phi$ and $K^{*0}$ decay branching fractions as well as corresponding
systematic uncertainties $\sigma$ on $\frac{f_s}{f_d}\frac{\mathcal{B}(\Bs\rightarrow \Jpsi \phi)}{\mathcal{B}(\Bd\rightarrow \Jpsi K^{*0})}$.}

\begin{tabular}{ l c c c}
\hline  \hline 
   Observable   & Value    &$ \sigma$   & Ref.  \\ \hline 
         $N_{B_s^0}$     &   $6640\pm 100 \pm 220   $                                    &    3.3\%    &       \\
         $N_{B_d^0}$      &  $36290\pm 320\pm 650    $                                  &    1.8\%      &      \\ \hline
         $\mathcal{R}_{\mathrm{eff}}$    &  $0.799\pm0.001\pm0.010     $                        &    1.3\%       &   \\  
         $\mathcal{B}(\phi \rightarrow K^{+}K^{-})$   &  $0.489 \pm0.005$           &    1.0\%     & \cite{PDG}        \\ 
         $\mathcal{B}( K^{*0}\rightarrow K^{+}\pi^{-})$ & $0.66503\pm0.00014 $         &    0.02\%      & \cite{PDG}       \\  \hline   
          Total  &  & 4.1\% & \\ \hline        

\hline  \hline 
\end{tabular}
\label{tab:systematic}
\end{table}

From the ratio $N_{B_s^0}/ N_{B_d^0}$ after efficiency correction and division by $\phi$ and $K^{*0}$ decay branching fractions,  ATLAS measures

\begin{equation}
\frac{f_s}{f_d}\frac{\mathcal{B}(\Bs\rightarrow \Jpsi \phi)}{\mathcal{B}(\Bd\rightarrow \Jpsi K^{*0})}=0.199 \pm 0.004(\mathrm{stat}) \pm 0.008(\mathrm{sys}).
\label{eq:fsfd2}  
\end{equation}

A perturbative QCD prediction~\cite{BsSM} yields

\begin{align*}
\frac{\mathcal{B}(\Bs\rightarrow \Jpsi \phi)}{\mathcal{B}(\Bd\rightarrow \Jpsi K^{*0})}=0.83^{+0.03}_{-0.02}(\omega_B) ^{+0.01}_{-0.00}(f_M)^{+0.01}_{-0.02}(a_i)^{+0.01}_{-0.02}(m_c),
\end{align*}

where the uncertainties result from the shape parameter $\omega_B$ of the $B$ meson wave function, meson decay constants $f_M$, Gegenbauer moments $a_i$ in the wave functions of the light vector mesons and the $c$-quark mass. Adding all contributions linearly yields a 7.1\% theory error. Using this prediction, the ratio of fragmentation fractions is measured to be
\begin{equation} 
\frac{f_s}{f_d}=0.240 \pm 0.004(\mathrm{stat}) \pm 0.010(\mathrm{sys}) \pm 0.017(\mathrm{th}).
\label{eq:fsfd3}  
\end{equation}

\begin{figure}
\centering
\includegraphics[width=3.0in]{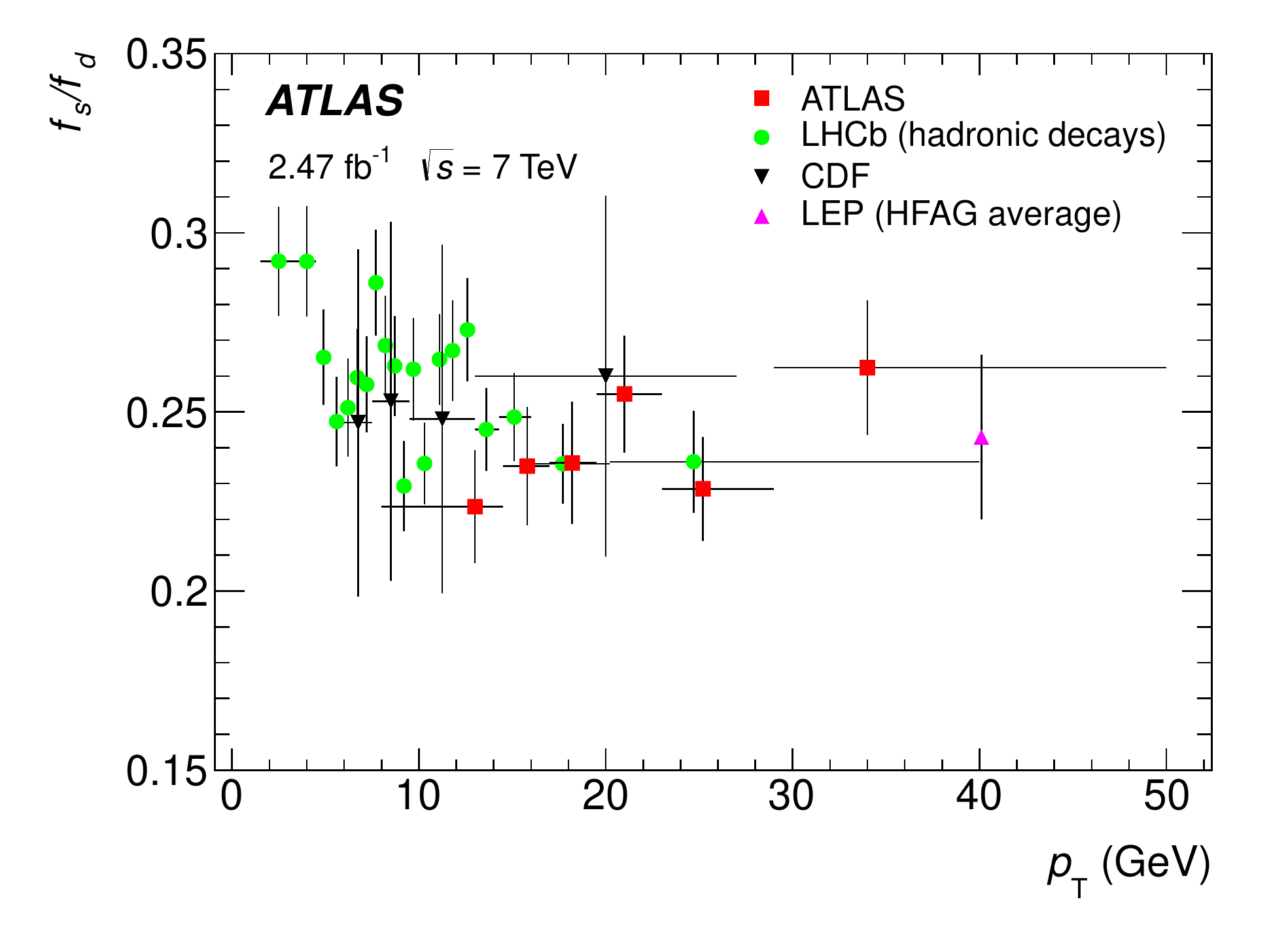}
\includegraphics[width=3.0in]{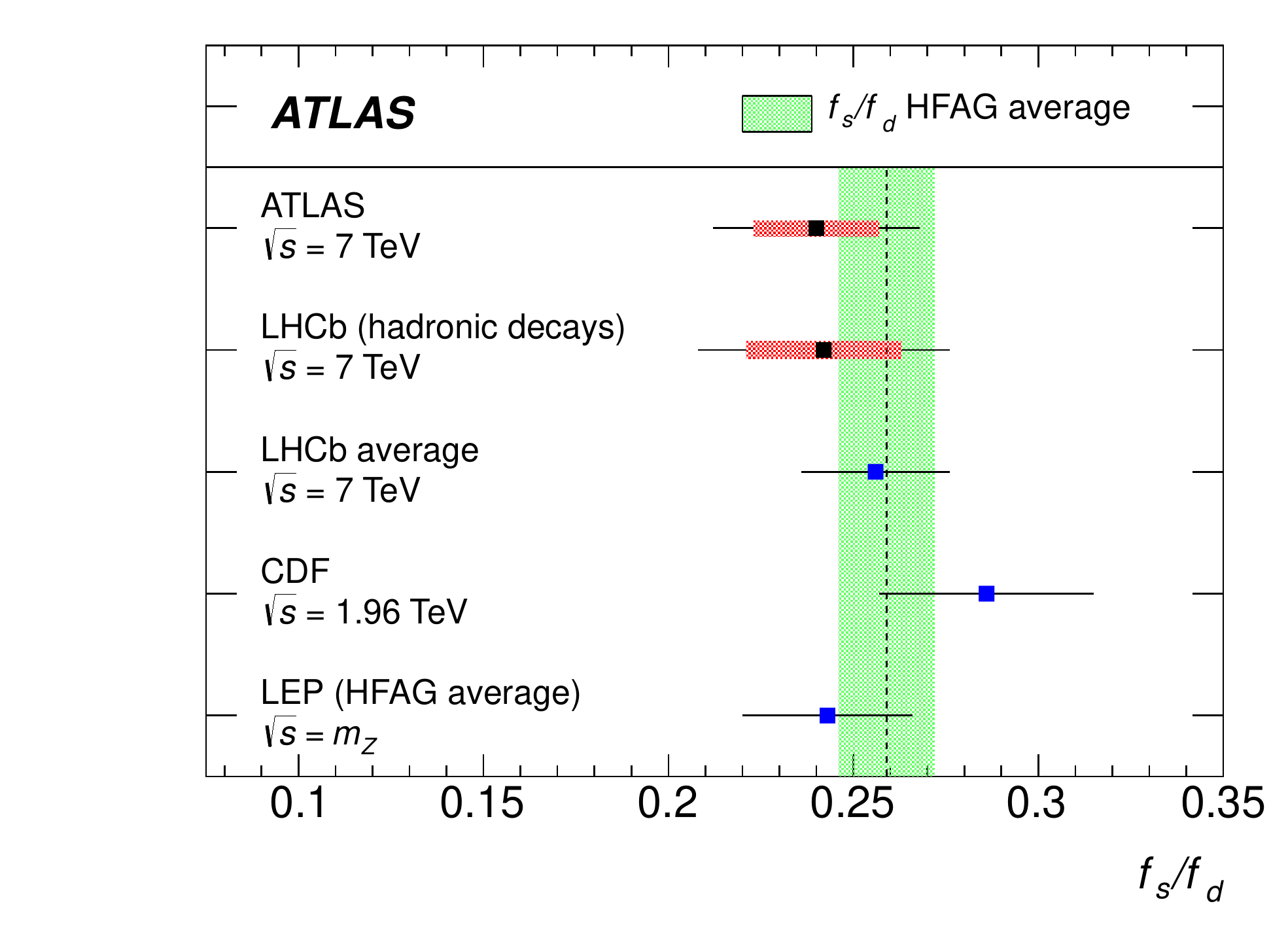}
\caption{(Left) Measurements of $f_s/f_d$ versus $B$ meson $\pt$ for CDF~\cite{CDF}, LHCb~\cite{LHCB_fsfd} and ATLAS, where the ATLAS data points are plotted at the average $\pt$ of the events in each bin. The error bars show statistical and systematic errors added in quadrature. The LEP ratio, taken from Ref.~\cite{HFAG}, is plotted at an average $\pt$ value in $Z$ decays. (Right) Measurements of $f_s/f_d$ (black and blue points with error bars) from LEP~\cite{HFAG}, CDF~\cite{HFAG}, LHCb~\cite{LHCB_fsfd, LHCbUpdated} and ATLAS. The total experimental error (thin black) is added linearly to the theory error (thick red). The green-shaded region shows the HFAG average obtained using the blue points.}
\label{fig:Pt}
\end{figure}

Figure~\ref{fig:Pt} (right) shows the ATLAS $f_s/f_d$ measurement in comparison with results from LEP~\cite{HFAG}, CDF~\cite{HFAG,CDF} and LHCb~\cite{LHCB_fsfd, LHCbUpdated}. The ratio $f_s/f_d$ may depend on $\pt$ and $\eta$ of the $B$ meson, e.g. LHCb observes a $\pt$ but no $\eta$ dependence of $f_s/f_d$~\cite{LHCB_fsfd}. Figure~\ref{fig:Pt} (left) shows the $\pt$ dependence of $f_s/f_d$ for ATLAS and that of other experiments. To investigate the $\pt$ and $\eta$ dependences of $f_s/f_d$, the data sample is divided into six  $\pt$ bins in the range 8 GeV $< \pt <$ 50 GeV and into four $\eta$ bins for $|\eta|< 2.5$ such that the number of events in each bin is approximately equal. The $f_s/f_d$ distributions as a function of $\pt$ and $\eta$ have been fitted with a uniform (first-order polynomial) distribution yielding fit $p$-values 0.54 (0.66) and 0.66 (0.49), respectively. No significant $f_s/f_d$ dependence on $\pt$ and $|\eta|$ is seen at the present level of accuracy. 

In summary, this Letter reports on the first ATLAS measurement of the ratio of $B^0_s \rightarrow J/\psi \phi$ and $B^0_d \rightarrow J/\psi K^{*0}$ branching fractions multiplied by the ratio of fragmentation fractions $f_s/f_d$ from which $f_s/f_d$ is determined.  The data were produced at the LHC in $pp$ collisions at $\sqrt{s}=7\ \mathrm{TeV}$ and correspond to an integrated luminosity of 2.47 $\rm fb^{-1}$. This $f_s/f_d$ measurement, obtained with a new approach, agrees with the LHCb~\cite{LHCB_fsfd, LHCbUpdated} results improving the world average considerably. A comparison with the CDF~\cite{HFAG,CDF} measurement and the LEP~\cite{HFAG} average confirms the universality of $f_s/f_d$. The ATLAS data show no dependence on $\pt$ nor on $|\eta|$ within the kinematic range tested.

\section*{Acknowledgements}

We thank CERN for the very successful operation of the LHC, as well as the
support staff from our institutions without whom ATLAS could not be
operated efficiently.

We acknowledge the support of ANPCyT, Argentina; YerPhI, Armenia; ARC,
Australia; BMWFW and FWF, Austria; ANAS, Azerbaijan; SSTC, Belarus; CNPq and FAPESP,
Brazil; NSERC, NRC and CFI, Canada; CERN; CONICYT, Chile; CAS, MOST and NSFC,
China; COLCIENCIAS, Colombia; MSMT CR, MPO CR and VSC CR, Czech Republic;
DNRF, DNSRC and Lundbeck Foundation, Denmark; EPLANET, ERC and NSRF, European Union;
IN2P3-CNRS, CEA-DSM/IRFU, France; GNSF, Georgia; BMBF, DFG, HGF, MPG and AvH
Foundation, Germany; GSRT and NSRF, Greece; ISF, MINERVA, GIF, I-CORE and Benoziyo Center,
Israel; INFN, Italy; MEXT and JSPS, Japan; CNRST, Morocco; FOM and NWO,
Netherlands; BRF and RCN, Norway; MNiSW and NCN, Poland; GRICES and FCT, Portugal; MNE/IFA, Romania; MES of Russia and ROSATOM, Russian Federation; JINR; MSTD,
Serbia; MSSR, Slovakia; ARRS and MIZ\v{S}, Slovenia; DST/NRF, South Africa;
MINECO, Spain; SRC and Wallenberg Foundation, Sweden; SER, SNSF and Cantons of
Bern and Geneva, Switzerland; NSC, Taiwan; TAEK, Turkey; STFC, the Royal
Society and Leverhulme Trust, United Kingdom; DOE and NSF, United States of
America.

The crucial computing support from all WLCG partners is acknowledged
gratefully, in particular from CERN and the ATLAS Tier-1 facilities at
TRIUMF (Canada), NDGF (Denmark, Norway, Sweden), CC-IN2P3 (France),
KIT/GridKA (Germany), INFN-CNAF (Italy), NL-T1 (Netherlands), PIC (Spain),
ASGC (Taiwan), RAL (UK) and BNL (USA) and in the Tier-2 facilities
worldwide.

\printbibliography

\newpage 
\begin{flushleft}
{\Large The ATLAS Collaboration}

\bigskip

G.~Aad$^{\rm 85}$,
B.~Abbott$^{\rm 113}$,
J.~Abdallah$^{\rm 151}$,
O.~Abdinov$^{\rm 11}$,
R.~Aben$^{\rm 107}$,
M.~Abolins$^{\rm 90}$,
O.S.~AbouZeid$^{\rm 158}$,
H.~Abramowicz$^{\rm 153}$,
H.~Abreu$^{\rm 152}$,
R.~Abreu$^{\rm 116}$
% ...
\bigskip
\\
% ...
$^{11}$ Institute of Physics, Azerbaijan Academy of Sciences, Baku, Azerbaijan\\
% ...
$^{85}$ CPPM, Aix-Marseille Universit{\'e} and CNRS/IN2P3, Marseille, France\\
% ...
$^{90}$ Department of Physics and Astronomy, Michigan State University, East Lansing MI, United States of America\\
% ...
$^{107}$ Nikhef National Institute for Subatomic Physics and University of Amsterdam, Amsterdam, Netherlands\\
% ...
$^{113}$ Homer L. Dodge Department of Physics and Astronomy, University of Oklahoma, Norman OK, United States of America\\
% ...
$^{116}$ Center for High Energy Physics, University of Oregon, Eugene OR, United States of America\\
% ...
$^{151}$ Institute of Physics, Academia Sinica, Taipei, Taiwan\\
$^{152}$ Department of Physics, Technion: Israel Institute of Technology, Haifa, Israel\\
$^{153}$ Raymond and Beverly Sackler School of Physics and Astronomy, Tel Aviv University, Tel Aviv, Israel\\
% ...
$^{158}$ Department of Physics, University of Toronto, Toronto ON, Canada\\
% ...
\end{flushleft}

\end{document}